\documentclass{tlp}
\usepackage{aopmath}
\usepackage{color}

\newtheorem{definition}{Definition} 
\newtheorem{example}{Example} 

\newtheorem{claim}{Claim}
\newenvironment{proofclaim}[1]{{\it Proof of Claim #1.\\}}{\hspace*{1em}\hbox{\proofbox}\endtrivlist}

\newcommand{\coNP}{\text{coNP}}
\newcommand{\NP}{\text{NP}}
\newcommand{\A}{\mbox{$\mathcal{A}$}}
\newcommand{\G}{\mbox{$\mathcal{G}$}}

\renewcommand{\P}{\mbox{$\mathcal{P}$}}
\newcommand\nbd{not~}

\title{On the complexity of identifying Head Elementary Set Free programs
\footnote{To appear in Theory and Practice of Logic Programming (TPLP).}}
\author[F. Fassetti and L. Palopoli]
{Fabio Fassetti$^{\text{\scriptsize\textdagger}}$\\
ICAR/CNR\\
via P. Bucci, 41C\\
87036, Rende (CS), Italy\\
E-mail: f.fassetti@deis.unical.it\\
\and
Luigi Palopoli\\
DEIS, University of Calabria\\
via P. Bucci, 41C\\
87036, Rende (CS), Italy\\
E-mail: palopoli@deis.unical.it
}

\submitted{11 December 2008}
\revised{29 April 2009}
\accepted{22 May 2009}

\begin{document}

\maketitle

\begin{abstract}
\renewcommand{\thefootnote}{\fnsymbol{footnote}}
\footnotetext[2]{This work was partly done while the author was affiliated with \emph{DEIS, University of Calabria}.}
Head-elementary-set-free programs were proposed in \cite{GebserLL07} and shown
to generalize over head-cycle-free programs while retaining their nice
properties. It was left as an open problem in \cite{GebserLL07} to establish
the complexity of identifying head-elementary-set-free programs. This note
solves the open problem, by showing that the problem is complete for $\coNP$.
\end{abstract}

\section{Introduction}
Disjunctive Logic Programming (DLP) is a highly declarative yet powerful
knowledge representation and problem solving formalism. However, the high
expressive power of DLP corresponds to a high complexity of the associated entailment
problems \cite{DantsinEGV01}. Therefore, the task of defining easily recognizable
fragments of DLP characterized by lower complexities than the general language
has been looked at as a relevant problem in the literature, since general DLP
resolution engines can speed up their computation by identifying subprograms
matching those definitions. For instance, the DLV engine \cite{LeonePFEGPS06} takes
advantage of identifying head-cycle-free (HCF) (sub)programs
\cite{Ben-EliyahuD94,Ben-Eliyahu-ZoharyP97} in resolving disjunctive logic
programs under the stable model semantics. Head-elementary-set-free (HEF)
programs were recently introduced in \cite{GebserLL07} as a strict
generalization of HCF programs featuring the same nice properties of that
smaller class. In detail, likewise HCF programs, HEF programs can be turned
into equivalent nondisjunctive programs in polynomial time and space by
shifting. As such, HEF programs can be regarded as ``easy'' disjunctive
programs, since they actually denote syntactic variants of nondisjunctive
coding. This fact has several formal consequences, which are precisely
accounted for in \cite{GebserLL07}. Just for an example, while checking for a
disjunctive program to have a stable model is $\Sigma^P_2$-complete in general,
it is $\NP$-complete for HEF programs.

It is therefore important to devise procedures to identify
head-elementary-set-free programs. However, while checking for a program to be
HCF can be done in linear time \cite{Ben-EliyahuD94}, the complexity of
identifying HEF programs is a problem left open in \cite{GebserLL07}, where it
is read that: {\em It is an open question whether identifying HEF programs is
tractable \dots}. This note is intended to solve such an open problem, by
showing that identifying HEF programs is, in fact, $\coNP$-complete. Therefore,
while HEF programs share several common properties with HCF programs, to
identify them is much more difficult from the computational complexity
standpoint.

The rest of the note is organized as follows. Preliminaries about DLP are
illustrated in the next section. Section \ref{HEF} recalls the definition of
HEF programs and provides a couple of preliminary results. Section
\ref{sect:complexityMemb} and Section \ref{sect:complexityHard} settle the
complexity of the problem accounting for the membership in $\coNP$ and its
$\coNP$-hardness, respectively.

\section{Preliminaries}\label{sect:prelim}

In this section we recall basic definitions about propositional disjunctive
logic programming.

A \emph{literal} is a propositional atom $a$ or its negation $\nbd a$. A rule
is an expression of the form $B,F \rightarrow H$, where $H$, $B$ and $F$ are
set of literals. In particular, sets $H$ and $B$ consist of positive atoms,
whereas $F$ consists of negated atoms. $H$ and $B\cup F$ are referred to as,
respectively, the head and body of the rule. If $|H|>1$ then the rule is called
\textit{disjunctive}, otherwise it is called \textit{non-disjunctive}.

A program $\P$ is a finite set of rules.
If there is some disjunctive rule in
$\P$ then $\P$ is called \textit{disjunctive}, otherwise it is called
\textit{non-disjunctive}. A set $S$ of atoms is called a \textit{disjunctive
set} for $\P$ if and only if there exists at least one rule $\delta: B,
F\rightarrow H$ in $\P$ such that $|H\cap S|>1$.

An interpretation $I$ of $\P$ is a  set of atoms from $\P$. An atom is {\em true}
in the interpretation $I$ if $a \in I$. A literal $\nbd a$ is {\em true} in $I$
if $a \not\in I$. A conjunction $C$ of  literals is true in $I$ if all the
literals in $C$ are true in $I$. A rule $B,F \rightarrow H$ is true in $I$ if
either $H$ is true in $I$ or $B \wedge F$ is false in $I$. An
interpretation $I$ is a model for a program $\P$ if all rules occurring in $\P$
are true in $I$. A model $M$ for $\P$ is minimal if no proper subset of $M$ is a
model for $\P$. A model $M$ of $\P$ is {\em stable} if $M$ is a minimal model of
the reduct of $\P$ w.r.t $M$, denoted by $\P^M$, that is the program built from
$\P$ by (1) removing all rules that contain a negative literal $\nbd a$ in the
body with $a \in M$, and (2) removing all negative literals from the remaining
rules \cite{GelfondL88}.

\begin{example}
Consider for example the following program:
\begin{eqnarray*}
\begin{array}{rrll}
\P = \{& a            & \rightarrow b, c & \\
       & \nbd a, d    & \rightarrow e & \\
       & c, \nbd b, f & \rightarrow e & \\
       & \nbd b       & \rightarrow a & \}
\end{array}
\end{eqnarray*}
and the interpretation $M=\{a, c\}$.
The ground positive program $\P^M$ is the following:
\begin{eqnarray*}
\begin{array}{rrll}
\P^M = \{& a    & \rightarrow b, c & \\
         & c, f & \rightarrow e & \\
         &      & \rightarrow a & \}
\end{array}
\end{eqnarray*}
Since $M$ is a minimal model of $\P^M$, $M$ is a stable model of $\P$.
\end{example}

\section{Head-elementary-set-free programs}
\label{HEF}

In this section, we recall the definition of HEF programs \citeS{GebserLL06} and provide a couple of preliminary results which will be useful in the following. We begin with introducing the concepts of outbound and elementary set.

\begin{definition}[Outbound Set\citeS{GebserLL06}]
Let $\P$ be a disjunctive program. For any set $Y$ of atoms occurring in $\P$, a subset $Z$ of $Y$ is \emph{outbound} in $Y$ for $\P$ if there is a rule $\delta: B, F\rightarrow H$ in $\P$ such that: (i) $H\cap Z\neq \emptyset$; (ii) $B\cap (Y\backslash Z)\neq \emptyset$; (iii) $B\cap Z = \emptyset$ and (iv) $H\cap (Y\backslash Z) = \emptyset$.
\end{definition}
Intuitively, $Z \subseteq Y$ is outbound in $Y$ for $\P$ if there exists a rule
$\delta$ in $\P$ such that the partition of $Y$ induced by $Z$ $(that is, \langle Z;
Y\setminus Z \rangle)$ separates head from body atoms of $\delta$.
\begin{example}\label{ex:elementary_set}
Consider, for example, the program
\begin{eqnarray*}
\begin{array}{rrll}
\P_{ex} = \{& a   & \rightarrow b, c & \\
            & c   & \rightarrow b    & \\
            & b   & \rightarrow c    & \\
            & b   & \rightarrow a    & \\
            & b,c & \rightarrow d    & \}
\end{array}
\end{eqnarray*}
and the set $E_{ex}=\{a,b,c\}$. Consider, now, the subset $O=\{a,b\}$ of $E_{ex}$. $O$ is outbound in $E_{ex}$ for $\P_{ex}$ because of the rule $c\rightarrow b$, since $c \in E_{ex}\setminus O$, $c\not\in O$, $b\in O$ and $b\not\in E_{ex}\setminus O$.
\end{example}
\begin{definition}[Elementary Set\citeS{GebserLL06}]
Let $\P$ be a disjunctive program. For any nonempty set $Y$ of atoms occurring
in $\P$, $Y$ is
\emph{elementary} for $\P$ if all nonempty proper subsets of $Y$
are outbound in $Y$ for $\P$.
\end{definition}
For example, the set $E_{ex}$ of Example \ref{ex:elementary_set} is elementary
for the
program $\P_{ex}$, since each nonempty proper subset of $E_{ex}$ is
outbound in $E_{ex}$ for $\P_{ex}$.

\begin{definition}[Head-Elementary-Set-Free Program\citeS{GebserLL07}]
Let $\P$ be a disjunctive program. $\P$ is \emph{Head Elementary Set Free
(HEF)} if for each
rule $B, F\rightarrow H$ in $\P$, there is no elementary set
$E$ for $\P$ such that $|E\cap H|>1$.
\end{definition}

So, a program $\P$ is HEF if there is no elementary set containing two or more
atoms all appearing in the head of one rule of $\P$.

For example, the program $\P_{ex}$ of Example \ref{ex:elementary_set} is not
HEF, because for the rule $\delta: a\rightarrow b,c$, and the elementary set
$E_{ex}$: the intersection between the head of $\delta$ and $E_{ex}=\{a,b,c\}$ is $\{b,c\}$.

It follows from the definition that a program $\P$ is not HEF if and only if
there exists a set $X$ of atoms of $\P$ such that $X$ is both a disjunctive set
and an elementary set for $\P$.

Next, two theorems which are needed to prove our
main results, given in the following sections, are proved. In particular, Theorem
\ref{theo:depGraphConnected} tells about the connectedness of the subgraph an
elementary set induces into a program positive dependency graph and actually immediately follows from \cite{GebserLL06}. Theorem \ref{theo:singleHeadSingleBody}, instead, tells that any atom that occurs in an elementary set must be ``justified'' by at least two rules, that atom being the only one in its elementary set occurring in the head of the first rule and in the body of the second rule, respectively. We begin by defining the concept of a positive dependency graph of a program.

A directed graph $\G$, called \emph{positive dependency graph}, can be
associated with a disjunctive program $\P$. Specifically, for each rule $B,F
\rightarrow H$ of $\P$, each atom appearing in $H$ or in $B$ is associated with
a node in $\G$, and there is a directed edge $(m,n)$ from a node $m$ to a node
$n$ if the atom associated with $m$ is in $B$, and the atom associated with $n$
is in $H$.

\begin{theorem}\label{theo:depGraphConnected}
Let $E$ be an elementary set for a program $\P$ and let $\G$ be the positive dependency graph associated with $\P$.
The subgraph induced by $E$ is strongly connected.
\end{theorem}
\begin{proof}
The proof is given by contraposition. Specifically, it is supposed that the
subgraph induced by $E$ is not strongly connected and it is derived that $E$ is
not elementary.

If the subgraph induced by $E$ is not strongly connected, then there exists some pair of node $m$ and $n$ such that $n$ is not reachable from $m$.
Then consider the set $E'\subset E$ of all the nodes reachable from $m$, and the set $E\setminus E'$.
Since $n$ is not reachable from $m$, $E\setminus E'$ is not empty, and then $E'$ is a proper subset of $E$.
Moreover, since reachability is a transitive relation, all the nodes in $E\setminus E'$ are not reachable from any node in $E'$.
By definition of dependency graph, it follows that there is no rule $B,F\rightarrow H$ in $\P$ such that $B\cap E' \neq \emptyset$ and $H\cap (E\setminus E') \neq \emptyset$. Then $E\setminus E'$ is not outbound and, as a consequence, $E$ is not elementary.
\end{proof}

\begin{theorem}\label{theo:singleHeadSingleBody}
Let $\P$ be a disjunctive program, let $E$ be an elementary set for $\P$ such
that $|E| > 1$ and let $a$ be an atom belonging to $E$. Then: (i) there exists
at least one rule $\delta_1: B,F\rightarrow H$, such that $a\not\in B$, $B\cap
E \neq \emptyset$ and $H\cap E=\{a\}$, and (ii) there exists at least one rule
$\delta_2: B,F\rightarrow H$, such that $a\not\in H$, $B\cap E=\{a\}$ and
$H\cap E \neq \emptyset$.
\end{theorem}
\begin{proof}
\begin{description}
\item[(i)] Consider the set $O=\{a\}$. If no rule $\delta_1: B,F\rightarrow
    H$, such that $a\not\in B$, $B\cap E \neq \emptyset$ and $H\cap
    E=\{a\}$, existed in $\P$, then $O$ would not be outbound. Since
    $O\subset E$, $E$ would not be elementary.
\item[(ii)] Consider the set $O=E\backslash \{a\}$. If no rule $\delta_2:
    B,F\rightarrow H$, such that $a\not\in H$, $B\cap E=\{a\}$ and $H\cap E
    \neq \emptyset$, existed in $\P$, then $O$ would not be outbound in $E$
    and then $E$ would not be elementary.
\end{description}
\end{proof}

Theorem \ref{theo:singleHeadSingleBody} closes the preliminary part of this
note. In the following Sections \ref{sect:complexityMemb} and
\ref{sect:complexityHard}, the complexity of identifying HEF programs is
analyzed.

\section{Complexity Analysis: Membership}\label{sect:complexityMemb}

In this section, the membership of the problem in the class $\coNP$ is proved.
To this end, some new properties of HEF programs are shown next.

Let $X$ be a set of atoms of a disjunctive logic program $\P$.
In the following,
$\P_X$ will denote the disjunctive logic program built as follows: for each
rule $\delta: B,F \rightarrow H$ of $\P$, add to $\P_X$ the rule $\delta':
B'\rightarrow H'$ obtained as the \emph{projection} of $\delta$ on $X$, namely
$B'$ is $B\cap X$ and $H'$ is $H\cap X$, if both $B'$ and $H'$ are not empty.

The following lemma is immediately proved.

\begin{lemma}\label{lemma:p_e}
Let $\P$ be a logic program.
$E$ is an elementary set for $\P$ if and only if $E$ is an elementary set for $\P_E$.
\end{lemma}

As a consequence of the above lemma, the definition of outbound set can be
rewritten as follows: \emph{let $\P$ be a disjunctive logic program, and let
$E$ be a set of atoms of $\P$. A subset $O$ of $E$ is outbound in $E$ for $\P$
if and only if there is a rule $\delta: B' \rightarrow H'$ in $\P_E$ such that
$\emptyset \subset H' \subseteq O$ and $\emptyset \subset B' \subseteq
E\backslash O$}.

The following lemma states that elementary sets of a program $\P$ are preserved
in supersets of $\P$.

\begin{lemma}\label{lemma:invariability}
Let $\P$ be a logic program, and $\P^{red}\subseteq \P$ a logic program consisting of a subset of the rules of $\P$.
If $E$ is an elementary set for $\P^{red}$, then $E$ is an elementary set for $\P$ as well.
\end{lemma}
\begin{proof}
If a set $E$ is an elementary set in $\P^{red}$ then, by definition, each
nonempty proper subset $S$ of $E$ is outbound in $E$ for $\P^{red}$ and,
therefore, there is a rule $\delta: B, F\rightarrow H$ in $\P^{red}$ such that
$H\cap S \neq \emptyset$, $B\cap (E\setminus S) \neq \emptyset$, $B\cap S =
\emptyset$ and $H\cap (E\setminus S) = \emptyset$.

Clear enough, if $\P^{red}\subseteq \P$ then $\delta$ is also in $\P$ and, as a
consequence, each subset of $E$ is outbound in $E$ also for $\P$.
\end{proof}

Let $\P$ be a logic program, and $E$ an elementary set for $\P$. In the
following, each program $\P^{red}_E \subseteq \P_E$ is called a \emph{witness}
of $E$ if $E$ is elementary in $\P^{red}_E$. Note, in particular, that $\P_E$
is a witness of $E$.

By Lemma \ref{lemma:invariability}, $\P^{red}_E$ shows that $E$ is elementary
for $\P_E$, and by
Lemma \ref{lemma:p_e} also for $\P$.

An important property of HEF programs is stated in the following theorem.

\begin{theorem}\label{theo:ND-core}
Let $\P$ be a disjunctive logic program. $\P$ is not HEF if and only if there
exists a pair $(E, \P^{red}_{E})$ such that $E$ is a disjunctive set for $\P$
and $\P^{red}_E$ is both a non-disjunctive program and a witness of $E$.
\end{theorem}

\begin{proof}
For one direction, note that if such a pair exists, then $E$ is a disjunctive
set for $\P$ and, since it has a witness, it is also an elementary set for $\P$
and, therefore, $\P$ is not HEF.

Now, consider the case in which $\P$ is not HEF.
In the following, it is proved that for each pair $(S, \P^{red}_S)$ such that
$S$ is a disjunctive set, and $\P^{red}_S$ is a disjunctive witness of $S$,
there exists a pair $(S', \P^{red}_{S'})$ such that $S'$ is a disjunctive set
and $\P^{red}_{S'}$ is a witness of $S'$, such that the number of disjunctive
rules in $\P^{red}_{S'}$ is strictly less than that of disjunctive rules
occurring in $\P^{red}_S$.

Note that this would conclude the proof, since it would inductively imply the
existence of a pair $({S}^\ast, \P^{red}_{{S}^\ast})$ such that ${S}^\ast$ is a
disjunctive
set, $\P^{red}_{{S}^\ast} \subseteq \P_{{S}^\ast}$ is a witness of
${S}^\ast$ with no disjunctive rules.

Let $(S, \P^{red}_S)$ be a pair such that $S$ is a disjunctive set, and
$\P^{red}_S$ is a witness of~$S$. Note that at least one of these pairs exists
since, by definition, for each non-HEF program, there exists an elementary set
$E$ and, by Lemma \ref{lemma:p_e}, a witness $\P_E$ of $E$ therefore exists as
well.
Assume that $\P^{red}_S$ is a disjunctive program. Then, at least one rule
$\delta^\ast: B\rightarrow H$, $|H|>1$ belongs to $\P^{red}_S$. Two cases are
possible: (i) $S$ is not an elementary set for
$\P^{red}_S\setminus\{\delta^\ast\}$; (ii) $S$ is an elementary set for
$\P^{red}_S\setminus\{\delta^\ast\}$.

\begin{description}
\item[(i)] Since $S$ is not elementary for $\P^{red}_S \setminus
    \{\delta^\ast\}$, then there exists at least one proper subset of $S$
    which is not outbound in $S$ for $\P^{red}_S \setminus
    \{\delta^\ast\}$. In particular, let $S'$ be a minimal subset of $S$
which is not outbound in $S$ for $\P^{red}_S \setminus \{\delta^\ast\}$.
Since $S'$ is outbound in $\P^{red}_S$, $\delta^\ast$ is such that
$H\subseteq S'$ and $B \subseteq S\setminus S'$, namely, $\delta^\ast$ is
needed to prove $S'$ to be outbound. It is worth noting that, because of
$\delta^\ast$, $S'$ is a disjunctive set for $\P$. Consider now each
nonempty proper subset $S''$ of $S'$. Note that one of such subsets exists,
since $S'$ contains at least all of the atoms belonging to the head of
$\delta^\ast$, and then its cardinality is greater than $1$.

Since $S'$ is a mimimal subset of $S$ which is not outbound in
$\P^{red}_S\setminus\{\delta^\ast\}$, $S''$ is outbound in
$\P^{red}_S\setminus\{\delta^\ast\}$. Therefore, there exists a rule
$\delta': B'\rightarrow H'$ in $\P^{red}_S\setminus \{\delta^\ast\}$, such
that $\emptyset \subset H' \subseteq S''$ and $\emptyset \subset
B'\subseteq S\setminus S''$.

Moreover, it must hold that $S' \cap B' \neq \emptyset$. Indeed, were $S'
\cap B' = \emptyset$ then $\delta': B' \rightarrow H'$ would be a rule such
that $\emptyset \subset H'\subseteq S'' \subset S'$ and $\emptyset \subset
B'\subseteq S\setminus S'$; hence, because of $\delta'$, $S'$ would be
outbound also in $\P^{red}_S \setminus \{\delta^\ast\}$, which does not
hold by hypothesis.

Consider, now, the program $\P^{red}_{S'}$ consisting of the projections of
the rules $\delta: B\rightarrow H$ of $\P^{red}_S$ such that $B\cap S' \neq
\emptyset$ and $H\cap S' \neq \emptyset$. Note that, as the rule
$\delta^\ast$ has the body contained in $S\setminus S'$, the projection of
$\delta^\ast$ is not added to $\P^{red}_{S'}$.

Since, as stated above, the set $S'$ is such that for each nonempty proper
subset $S'' \subset S'$ there is a rule $\delta':B' \rightarrow H'$ in
$\P^{red}_{S}$ where $\emptyset \subset H'\subseteq S''$ and $\emptyset
\subset B' \subseteq S'\setminus S''$, it follows that $\delta'$ is also in
$\P^{red}_{S'}$ and, therefore,  $S''$ is outbound in $S'$; this implies,
in turn, that $\P^{red}_{S'}$ is a witness of $S'$.

Summarizing, for each pair $(S, \P^{red}_{S})$ such that $S$ is an
elementary set for $\P$ and $\P^{red}_{S}$ is a witness of $S$ containing
at least one disjunctive rule $\delta$, there exist both a non-empty
disjunctive set $S' \subset S$ such that $S'$ is a disjunctive set for $\P$
and a witness $\P^{red}_{S'}$ of $S'$, such that $\P^{red}_{S'}$ contains a
number of disjunctive rules strictly less than the number of disjunctive
rules occurring in $\P^{red}_S$ (as the former does not contain
$\delta^\ast$).

\item[(ii)] In this second case, consider the pair $(S', \P^{red}_{S'})$,
    where $S' = S$ and $\P^{red}_{S'} =\P^{red}_S
    \setminus\{\delta^\ast\}$. $S'$ is a disjunctive set for $\P$ and
    $\P^{red}_{S'}$ is a witness of $S'$ that does not contain the
    disjunctive rule $\delta^\ast$.

\end{description}
\end{proof}

\begin{example}\label{ex:membership_proof}
In order to clarify the proof of the Theorem \ref{theo:ND-core}, consider the following example.
Let $\P$ be the following program
\begin{eqnarray*}
\begin{array}{rrllcrrlll}
\P = \{& a   & \rightarrow b, c &   & \quad &\P^{red}_{S'} = \{& c & \rightarrow b &    \\
       & c   & \rightarrow b    &   & \quad &                  & b & \rightarrow e &    \\
       & b   & \rightarrow d    &   & \quad &                  & e & \rightarrow f &    \\
       & b   & \rightarrow e    &   & \quad &                  & f & \rightarrow e &    \\
       & d,e & \rightarrow f    &   & \quad &                  & e & \rightarrow c & \} \\
       & f   & \rightarrow e    &   & \quad &                  &   &               &    \\
       & e   & \rightarrow c    &   & \quad &                  &   &               &    \\
       & d   & \rightarrow a    &\} & \quad &                  &   &               &
\end{array}
\end{eqnarray*}
which is not HEF, since the set $E=\{a,b,c,d,e,f\}$ is elementary for $\P$.
Furthermore, $E$ is a disjunctive set, due to the rule $\delta^\ast:
a\rightarrow b,c$ and $\P$ is a witness of $E$. $E$ is not elementary for $\P
\setminus\{\delta^\ast\}$ since $S'=\{b,c,e,f\}$ is not outbound in $E$ for
$\P \setminus\{\delta^\ast\}$ and, moreover, $S'$ is a minimal non-outbound
subset of $E$. Note that $S'$ is outbound in $\P$ just for the presence of
$\delta^\ast$, and $S'$ is a disjunctive set since it contains the whole head
of $\delta^\ast$. Consider the program $\P^{red}_{S'}$. Since $S'$ is a minimal
non-outbound subset of $E$, each nonempty subset of $S'$ is outbound in
$\P^{red}_{S'}$, and then $S'$ is elementary for $\P^{red}_{S'}$. Summarizing,
$S'$ is a disjuctive set and is also an elemetary set for $\P^{red}_{S'}$ and
then for $\P$. Thus, $\P^{red}_{S'}$ is a witness of $S'$ and it is also
non-disjunctive, since it does not contain $\delta^\ast$.
\end{example}

Using the result stated in Theorem \ref{theo:ND-core}, it is possible to prove the $\coNP$-membership theorem.

\begin{theorem}[HEF Problem-Membership]
Let $\P$ be a disjunctive logic program. Deciding if $\P$ is HEF is in coNP.
\end{theorem}
\begin{proof}
By Therorem \ref{theo:ND-core}, a nondeterministic polynomial-time Turing
machine can disqualify the HEF-Problem by first guessing a pair $(Y,
\P^{red}_Y)$ where $Y$ is a set of atoms and $\P^{red}_Y$ is a non-disjunctive
program.
Next, the machine verifies in polynomial time that at least two atoms,
belonging to the head of a rule in $\P$, are contained in $Y$ (that is, that
$Y$ is a disjunctive set for $\P$) and, finally, checks that $Y$ is an
elementary set for $\P^{red}_Y$, by verifying that $\P^{red}_Y$ is a witness of
$Y$. This last task can be accomplished in polynomial time as stated in
\cite{GebserLL06}. If this holds, by Lemmata \ref{lemma:p_e} and
\ref{lemma:invariability}, it follows that $Y$ is elementary for $\P$ and then
$\P$ is not HEF.
\end{proof}

\section{Complexity Analysis: Hardness}\label{sect:complexityHard}

In this section the $\coNP$-hardness of the problem is proved.

Let $\Phi=C_1\wedge \dots \wedge C_n, n\ge 1$ be a $3$-CNF formula, namely a
conjunctive Boolean formula where each clause $C_i$ consists exactly of three
literals. From $\Phi$, a logic program $\P^\Phi$ is constructed as follows. Let
$A_1,\dots,A_m$ be the variables of $\Phi$; and let $\A^\Phi$ be a set of atoms
consisting of: an atom $\phi$; an atom $a_i$ and an atom $n a_i$ for each
variable $A_i$; an atom $c_i$ for each clause $C_i$; and, finally, two further
atoms $c_0$ and $c_{n+1}$. Thus, note that $\A^\Phi$ is always non-empty.
In the following, the atom $n a_i$ is referred to as
the \emph{opposite} of the atom $a_i$ and vice versa. For each atom $c_i$,
$V(c_i)$ denotes the set of atoms associated with the literals appearing in the
clause $C_i$. In particular, an atom $a_j$ belongs to $V(c_i)$ if $A_j$ appears
in $C_i$ and $n a_j$ belongs to $V(c_i)$ if $\neg A_j$ appears in
$C_i$. Moreover, for each atom $c_i$, $N{V}(c_i)$ denotes the set of the
opposites of the atoms in $V(c_i)$, namely the atom $a_j$ (resp. $n a_j$) is in
$N{V}(c_i)$ if $n a_j$ (resp. $a_j$) is in $V(c_i)$.
$\P^\Phi$, the disjunctive program associated with $\Phi$ and built on $\A^\Phi$, consists in the following rules:
\begin{enumerate}
\item $\phi \rightarrow c_0 \vee c_{n+1}$
\item $c_0 \rightarrow c_1$
\item $c_i \wedge \alpha^i_j \rightarrow c_{i+1}$, for each $1\le i \le n$ and for each $\alpha^i_j\in N{V}(c_i), 1\le j \le 3$
\item $c_{n+1} \wedge n a_1 \rightarrow a_1$
\item $c_{n+1} \wedge a_1 \rightarrow n a_1$
\item $a_i \wedge n a_{i+1} \rightarrow a_{i+1}$, $1\le i \le m-1$;
\item $a_i \wedge a_{i+1} \rightarrow n a_{i+1}$, $1\le i \le m-1$;
\item $n a_i \wedge n a_{i+1} \rightarrow a_{i+1}$, $1\le i \le m-1$;
\item $n a_i \wedge a_{i+1} \rightarrow n a_{i+1}$, $1\le i \le m-1$;
\item $a_m \wedge n a_m \rightarrow c_0$;
\end{enumerate}

\begin{theorem}[HEF Problem-Hardness]
Let $\P$ be a disjunctive logic program. Deciding if $\P$ is HEF is coNP-hard.
\end{theorem}
\begin{proof}

The proof is given by reduction of $3$-SAT, which is well known to be $\NP$-complete \cite{GareyJ79}.

Let $\Phi=C_1\wedge \dots \wedge C_n$ be a $3$-CNF and $\P^\Phi$ the
disjunctive program associated with $\Phi$. First, we note that the size of
$\P^\Phi$ is polynomially bounded in the size of $\Phi$. Next, it is proved
that $\P^\Phi$ is not HEF if and only if $\Phi$ is satisfiable.

Since the only rule of $\P^\Phi$ containing more than one atom in the head is
$\phi\rightarrow  c_0 \vee c_{n+1}$, in order to prove that $\P^\Phi$ is not HEF, an elementary set $E$ containing both $c_0$ and
$c_{n+1}$ must be found.

Before proceeding with the proof of the theorem, some claims are shown about this.

\begin{claim}\label{claim1}
$E$ does not contain both $a_i$ and $n a_i$ for any $i\in [1,m]$.
\end{claim}
\begin{proofclaim}{\ref{claim1}}
If there existed $i$ such that both $a_i$ and $n a_i$ are in $E$, then the set
$\{a_i, n a_i\}\subset E$ would not be outbound in $E$ and $E$ would not be
elementary.
\end{proofclaim}

\begin{claim}\label{claim2}
$E$ contains $c_j$, for all $1\le j\le n$.
\end{claim}
\begin{proofclaim}{\ref{claim2}}
Because of Theorem \ref{theo:depGraphConnected}, the subgraph induced by the
atoms in $E$ must be strongly connected; then, since $E$ contains both $c_0$
and $c_{n+1}$ and since the only path from $c_0$ to $c_{n+1}$ passes through
atoms $c_1,\dots,c_n$, all these atoms must belong to $E$.
\end{proofclaim}

\begin{claim}\label{claim3}
$E$ contains at least one atom out of $a_i$ and $n a_i$, for each $i\in [1,m]$.
\end{claim}
\begin{proofclaim}{\ref{claim3}}
Because of Theorem \ref{theo:depGraphConnected}, the subgraph induced by the
atoms in $E$ must be strongly connected; then, since $E$ contains both $c_0$
and $c_{n+1}$ and since all the paths from $c_{n+1}$ to $c_0$ pass through
either the atom $a_i$ or the atom $n a_i$ for each $i\in [1,m]$, either the
atom $a_i$ or the atom $n a_i$ must belong to $E$.
\end{proofclaim}

\noindent Summarizing the results of previous claims, a potential elementary
set $E$ for $\P^\Phi$ consists of:

\begin{itemize}
\item the atoms $c_0,c_1\dots,c_n,c_{n+1}$;
\item either the atom $a_i$ or the atom $n a_i$ (but not both of them), for each $i\in[1,m]$.
\end{itemize}

\begin{claim}\label{claim:necCond}
Let $E$ be as described above. Then, for each clause $C_i$, at least one atom in $N{V}(c_i)$ is not in $E$.
\end{claim}
\begin{proofclaim}{\ref{claim:necCond}}
There are only three rules having $c_i$ in their body, namely $c_i \wedge \alpha^i_j\rightarrow
c_{i+1}$ for each $\alpha^i_j\in N{V}(c_i)$.
Due to Theorem \ref{theo:singleHeadSingleBody}, in order for $E$ to be elementary, at least one rule $B\rightarrow H$ such that $B\cap E=\{c_i\}$ must occur in $\P^\Phi$; then at least one atom $\alpha^i_j \in N{V}(c_i)$ has not to belong in $E$.
\end{proofclaim}

The above claim asserts that, in order for $E$ to be elementary, for each
clause $C_i$ a \emph{necessary} condition is that at least one atom in
$N{V}(c_i)$ must be not in $E$.
It can be shown that this is also a \emph{sufficient} condition.

\begin{claim}\label{claim:suffCond}
Let $E$ be as described above. Then, if for each clause $C_i$ at least one atom
in $N{V}(c_i)$ is not in $E$, then $E$ is an elementary set for $\P^\Phi$.
\end{claim}
\begin{proofclaim}{\ref{claim:suffCond}}
The proof is given by picking a generic nonempty proper subset $O$ of $E$ and
by showing that it is outbound in $E$ for $\P^\Phi$.

Let $Q\subset E$ be the subset of $E$ consisting of exactly one of the atoms
$a_i$ and $n a_i$ for each $i\in[1,m]$; and let $Q_i$ be the atom $a_i$ (resp.,
$n a_i$), if $a_i$ (resp., $n a_i$) belongs to $Q$. Moreover, let $\G_E$ denote
the subgraph induced by the atoms in $E$ and consider the path $\pi$ in $\G_E$
consisting of: (i) the directed edge from the $c_i$ to $c_{i+1}$ for each $0\le
i \le n$, (ii) the directed edge from $c_{n+1}$ to $Q_1$, (iii) the directed
edge from $Q_i$ to $Q_{i+1}$ for each $1\le i \le m-1$ and, finally (iv) the
directed edge from $Q_m$ to $c_0$. Note that $\pi$ is an Hamiltonian cycle.
Since $O$ is a nonempty proper subset of $E$ then at least one node of $E$ is
not in $O$. Therefore, there exists a pair of nodes $n_1$ and $n_2$ in $\G_E$
such that the atom $x_1$ associated with $n_1$ is in $E\setminus O$, the atom
$x_2$ associated with $n_2$ is in $O$ and there exists a directed edge from
$n_1$ to $n_2$ in $\pi$.
Since there exists a directed edge from $n_1$ to $n_2$, then there is a rule
$\delta: B\rightarrow H$ in $\P^\Phi$ such that $x_1\in B\cap E$ and $x_2\in
H\cap E$. In particular, it will be shown  next that there exists a rule
$\delta': B'\rightarrow H'$ such that $B'\cap E=\{x_1\}$ and $H'\cap
E=\{x_2\}$. Note that this will conclude the proof, since $O$ is outbound just
by the virtue of $\delta'$.

Since there exists a directed edge from $n_1$ to $n_2$, simply consider all the
pairs of atoms associated with the directed edges in $\pi$; the following cases
exhaust all possibilities: (i) $x_1 = c_i$ and $x_2 = c_{i+1}$ for some $0\le i
\le n$; (ii) $x_1=c_{n+1}$ and $x_2=Q_1$; (iii) $x_1=Q_i$ and $x_2=Q_{i+1}$ for
some $1\le i \le m-1$; (iv) $x_1=Q_m$ and $x_2=c_0$.

Consider case (i). Since for each clause $C_i$ at least one atom in $N{V}(c_i)$
is not in $E$, there exists at least one rule $\delta': c_i \wedge \alpha^i_j
\rightarrow c_{i+1}$ in $\P^\Phi$ such that the intersection between $E$ and
the body of $\delta$ is $\{c_i\}$.
As for case (ii), assume w.l.o.g. that $Q_1=a_1$ and then that $n a_1\not \in
E$. Then, the rule $\delta': c_{n+1} \wedge n a_1\rightarrow a_1$ is such that
the intersection between $E$ and the body of $\delta'$ is $\{c_{n+1}\}$.
Consider case (iii), assume w.l.o.g., that $Q_i=a_i$ and $Q_{i+1}=a_{i+1}$.
Then, the rule $\delta': a_i \wedge n a_{i+1} \rightarrow a_{i+1}$ is such that
the intersection between $E$ and the body of $\delta'$ is $\{a_i\}$.
Finally, as for case (iv), assume w.l.o.g., that $Q_m=a_m$. The rule $\delta':
a_m \rightarrow c_0$ is such that the intersection between $E$ and the body of
$\delta'$ is $\{a_m\}$.
\end{proofclaim}

\medskip
Now, the proof of the theorem can be resumed.

Let $X$ be a truth assignment to the variables in $\Phi$. Let $Q^X$ be the set
of atoms associated with $X$. In particular, $a_i$ (resp., $n a_i$) is in
$Q^X$, if $A_i$ is true (resp., false) in~$X$. It is proved that: $X$ is
satisfies $\Phi$, if and only if the set $E=\{c_0,\dots,c_{n+1}\}\cup Q^X$ is
elementary for $\P^\Phi$.
Note that this will conclude the theorem proof, since $E$ contains both $c_0$ and $c_{n+1}$.

\begin{description}
\item[($\Rightarrow$)] If $X$ satisfies $\Phi$ then $Q^X$ contains at least
    one atom $\alpha\in V(c_i)$ for each $c_i, i\in[1,n]$. Therefore, at
    least one atom, in particular the opposite of the atom $\alpha$, that
    belongs to $N{V}(c_i)$ for each $c_i, i\in[1,n]$, is not in $E$. Thus,
    by Claim \ref{claim:suffCond}, $E$ is elementary.
\item[($\Leftarrow$)] By Claim \ref{claim:necCond}, if $E$ is elementary
    then $Q^X$ does not contain any $\alpha\in N{V}(c_i)$ for each
    $c_i, i\in[1,n]$. Then, for each clause $C_i$, $Q^X$ contains one of
    the atoms associated with the literals satisfying $C_i$. Therefore, the
    truth assignment associated with $Q^X$ satisfies $\Phi$.
\end{description}

\end{proof}

\section{Conclusions}\label{sect:conclusions}
In this work the complexity of verifying if a disjunctive logic program is
head-elementary-set-free is analyzed. We have proved here that the
problem at hand is $\coNP$-complete, hereby providing an answer to a question
left open in \cite{GebserLL07}. This, basically negative, result leaves open
the further problem of singling out a polynomial-time recognizable fragment of
DLP, generalizing over HCF programs, while sharing their nice computational
characteristics. In this respect, a direction to go is supposedly that of
identifying some simple subclasses of programs for which checking for
head-elementary-set-freeness is easier than for the general case\footnote{Authors thank
one of the anonymous referees for having pointed this out.}.
\bibliographystyle{acmtrans}
\bibliography{hef}

\end{document}